\documentclass{PoS}

\title{Differential cross sections for top pair production at the LHC}

\ShortTitle{Top-quark pair production at the LHC}

\author{\speaker{Marco Guzzi},$^a$ Katerina Lipka$^{a}$ 
and Sven-Olaf Moch $^{b,c}$ \\
\llap{$^a$}Deutsches Elektronen-Synchrotron DESY\\
Notkestrasse 85, 22607 Hamburg, Germany\\
\llap{$^b$}Institut f\"ur Theoretische Physik, Universit\"at Hamburg, Luruper Chaussee 149
Hamburg, D-22761, Germany\\
\llap{$^c$}Deutsches Elektronen-Synchrotron DESY\\
Platanenallee 6, D-15738 Zeuthen, Germany\\
E-mail: \email{marco.guzzi@desy.de},\email{katerina.lipka@desy.de}, 
\email{sven-olaf.moch@desy.de}}

\abstract{
We present results of phenomenological studies for top-quark pair production at the LHC 
at the center of mass energy $\sqrt{S} = 7$ TeV.
The transverse momentum and rapidity distributions for final-state top quarks are 
calculated in perturbative QCD at approximate next-to-next-to-leading order ${\cal O}(\alpha_s^4)$ 
by using methods of threshold resummation beyond the leading logarithmic accuracy.
The theoretical predictions are obtained by using 
the computer code \textsc{DiffTop} and are compared to 
recent measurements by the ATLAS and CMS collaborations.
\textsc{Difftop} can be employed in the general 
case of heavy-quark pair production at hadron-hadron colliders and provides a basis 
for applications in QCD analyses for parton distribution functions determination.
}
 
\FullConference{XXII International Workshop on Deep-Inelastic Scattering and Related Subjects\\
                  28 April - 2 May 2014, Warsaw, Poland\\ 
                 {\tt Preprint DESY 14-155} }

\begin{document}

{\bf Introduction.}
Since its discovery at the Tevatron, the top quark has been playing 
an extremely important role in particle phenomenology. 

Its mass, a fundamental parameter of the Standard Model (SM),
is the largest in the quark families and it is close to 
the electroweak symmetry breaking (EWSB) scale. 
Therefore, the top quark behaves differently with respect to the other quarks for many reasons.
The leading decay channel of the top-quark into a $b$ quark and a $W$ boson, is  
mainly controlled by weak interactions, thus decay properties like
spin correlations and helicity can be investigated in a clean way in the decay products, before that hadronization takes place.
This is crucial for precision measurements and tests of the electroweak (EW) sector.
Furthermore, the mass of the top quark recently obtained from the combined results of the measurements 
of the CMS and ATLAS~\cite{ATLAS:2014wva} collaborations at the LHC,
is $m_t=173.3\pm 0.76$ GeV and since it is close to the mass of the Higgs boson, 
it gives us the possibility of studying the interplay 
between the Higgs sector and top-quark physics.

Due to the large mass, processes involving final-state top quarks in high-energy hadronic reactions
are excellent candidates to probe parton distribution functions (PDFs) of the proton in 
kinematic regions where these (particularly the gluon) are currently poorly constrained and
are correlated with the strong coupling constant $\alpha_s$ and top-quark mass $m_t$.

LHC run-I provided us with the possibility to perform precise measurements 
of total and differential cross sections for top-quark pair production at center-of-mass energies $\sqrt{S}=7$ and 8 TeV, 
recently published by the CMS~\cite{Chatrchyan:2012saa,Chatrchyan:2013faa} 
and ATLAS~\cite{Aad:2012hg,TheATLAScollaboration:2013eja,TheATLAScollaboration:2013dja} 
collaborations. These measurements are being used in multiple phenomenological analyses 
where $t\bar{t}$ data are used to test the
properties of the Standard Model (SM), QCD factorization and
to investigate possible signals of physics beyond the SM (BSM).

Total and differential cross sections for $t\bar{t}$ pair production 
at the LHC are mostly driven by the gluon-gluon luminosity, in which the gluon PDF 
is probed at large values of the parton momentum fraction $x\approx 0.1$.
The inclusion of $t\bar{t}$ pair production measurements in global QCD analyses 
for PDF determinations will allows us to investigate 
the correlation between the top-quark mass $m_t$, 
strong coupling constant $\alpha_s$, and the gluon.

In this brief paper we illustrate phenomenological results, documented in Refs.~\cite{Guzzi:2014wia,Guzzi:2013noa}, 
which are of importance for analyses at the LHC and are obtained by using the \textsc{DiffTop} code, which
provides a basis for applications in QCD analyses to determine PDFs and it will be soon released for public use.

\begin{figure}[ht]
\begin{centering}
\includegraphics[width=6cm, angle=0]{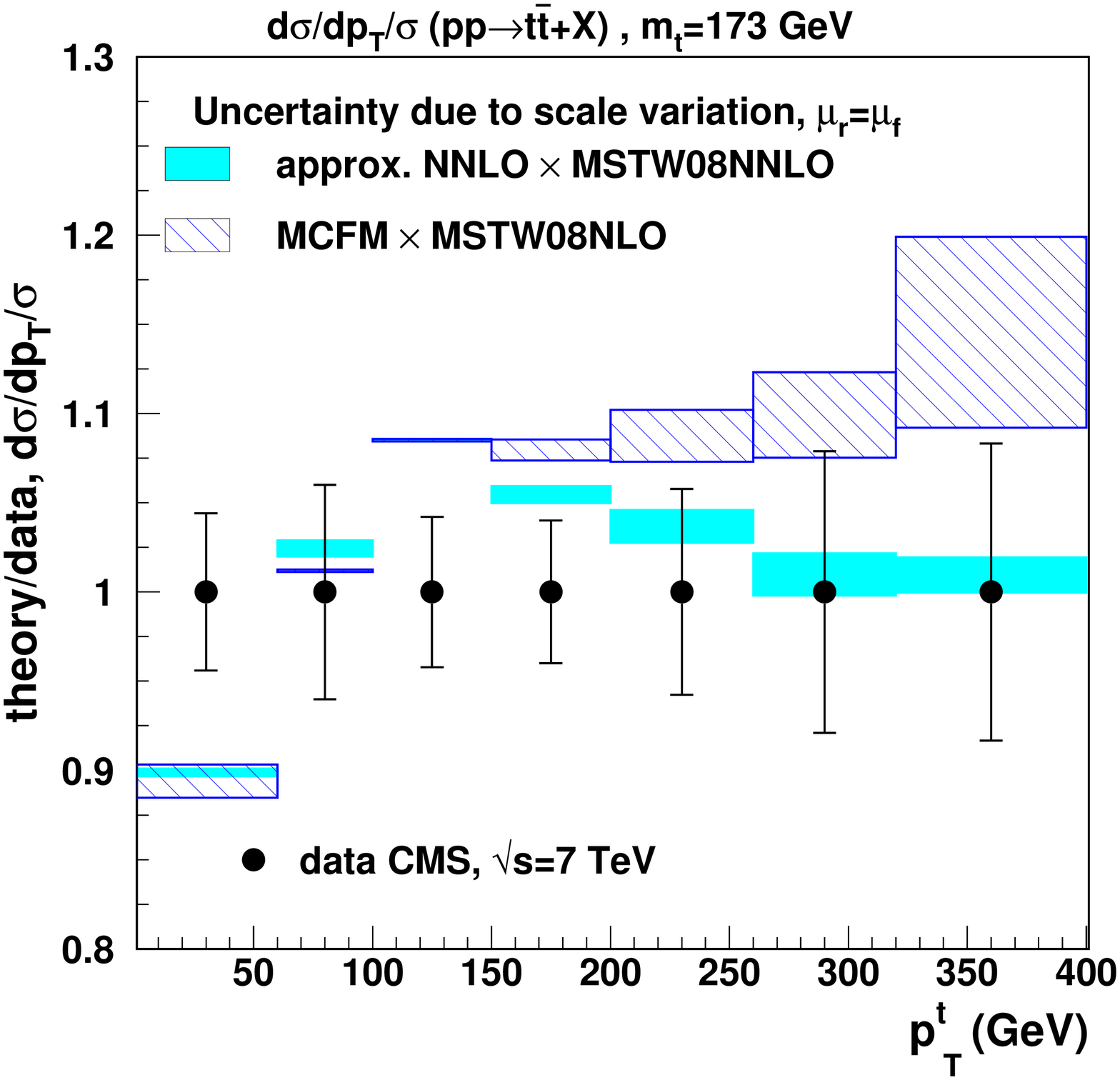}
\includegraphics[width=6cm, angle=0]{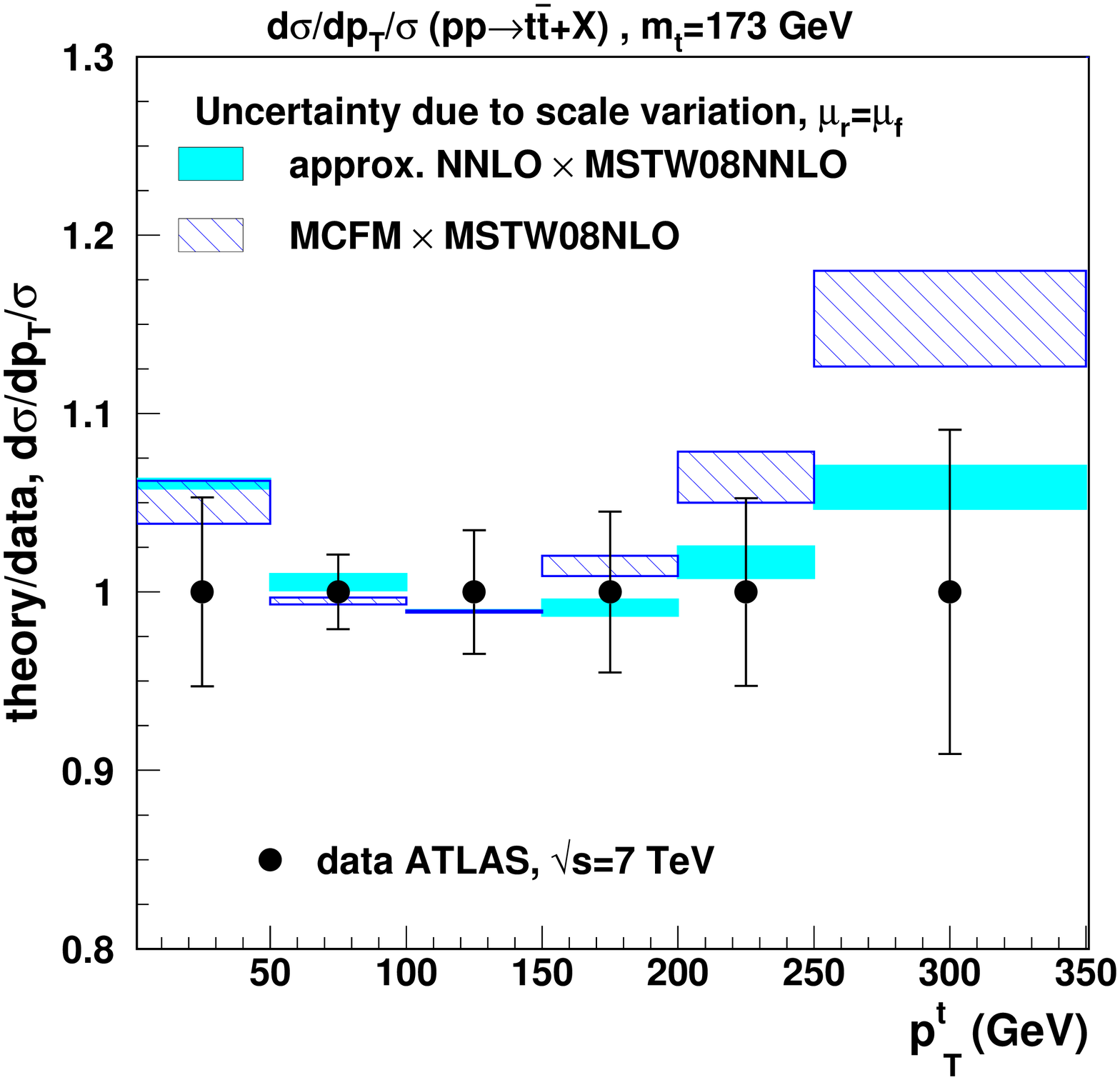}
\par\end{centering}
\caption{Study of scale uncertainties for \textsc{MCFM} and \textsc{DiffTop} calculations 
for the top-quark $p^t_T$ distribution. 
Ratio of theory over data for CMS (left)~\cite{Chatrchyan:2012saa} and ATLAS (right)~\cite{Aad:2012hg} 
measurements. Here MSTW08 NLO (NNLO) PDFs are used for the \textsc{MCFM} (\textsc{DiffTop}) calculation.
Renormalization and factorization scales are set to $\mu_R=\mu_F=m_t$ and varied such as $m_t/2\leq \mu_R=\mu_F\leq 2m_t$.
\label{mstw08-nlo-scale-unc}}
\end{figure}

{\bf The need for precision.}
To fully exploit the constraining power of the current data,
precise theoretical predictions are needed at the highest perturbative order possible, 
in which systematic uncertainties associated with renormalization/factorization ($\mu_R,\mu_F$) and other scales 
are reduced.
Efficient tools for the analyses, incorporating the current state-of-the-art of QCD calculations 
for $t\bar{t}$ observables are therefore necessary.
The QCD corrections to heavy-quark production at hadron colliders at the next-to-leading order 
(NLO), ${\cal O}(\alpha_s^3)$, are known since 
many years~\cite{Nason:1987xz,Nason:1989zy,Beenakker:1988bq,Meng:1989rp,Beenakker:1990maa,Mangano:1991jk}. 
The full calculation at next-to-next-to-leading order (NNLO), ${\cal O}(\alpha_s^4)$, 
for the inclusive cross section has been accomplished only recently~\cite{Czakon:2013goa,Czakon:2012pz,Czakon:2012zr,Baernreuther:2012ws} 
and required continuous efforts of the QCD community in calculating radiative corrections 
and in the development of computational tools~\cite{Czakon:2007wk,Czakon:2007ej,Mitov:2006xs,Ferroglia:2009ep,Ferroglia:2009ii,Czakon:2010td,Bierenbaum:2011gg,Baernreuther:2013caa}. 
The NNLO calculation for the inclusive cross section for the $t\bar{t}$ production is implemented 
in the \textsc{C++} computer programs \textsc{Top++}~\cite{Czakon:2011xx} and \textsc{Hathor}~\cite{Aliev:2010zk}.
The exact NLO calculations for $t\bar{t}$ total and differential cross sections are efficiently implemented 
into Monte Carlo (MC) numerical codes \textsc{MCFM}~\cite{Campbell:2000bg},
\textsc{MC@NLO}~\cite{Frixione:2003ei}, \textsc{aMCfast}~\cite{Bertone:2014zva}, \textsc{POWHEG}~\cite{Alioli:2010xa}, 
\textsc{MadGraph/MadEvent}~\cite{Alwall:2007st,Frederix:2009yq}. 
On the other hand, the NNLO corrections for $t\bar{t}$ differential cross sections are not yet available 
and NLO predictions seem to be not accurate enough to describe the current LHC data, because 
perturbative corrections are large and systematic 
uncertainties associated to various scales entering the calculation are important (see Fig.~\ref{mstw08-nlo-scale-unc}).
For this specific purpose, techniques of QCD threshold resummation provide us with 
theoretical tools to estimate the importance of perturbative higher orders in cross-section calculations~\cite{Sterman:1986aj,Catani:1989ne,Catani:1990rp,Kidonakis:1997gm,Laenen:1998qw,Bonciani:1998vc}.
By using threshold resummation methods one can derive approximate formulas 
beyond the NLO approximation, in which cross sections
are expanded in terms of the logarithmic enhanced contributions
and can therefore be written at various degrees of logarithmic accuracy.

\textsc{DiffTop} is a Mellin-space resummation computer code for calculating total and differential 
cross section for heavy-flavor production at hadron colliders at approximate NNLO ${\cal O}(\alpha_s^4)$.
It uses techniques of logarithmic expansion beyond the leading logarithmic accuracy in QCD threshold resummation
and the implementation strictly follows the derivation in Ref.~\cite{Kidonakis:2001nj} and references therein.
Other particulars of the calculation can be found in Refs.~\cite{Guzzi:2014wia,Kidonakis:2000ui,Kidonakis:2003tx,Kidonakis:2005kz,Kidonakis:2013zqa}.

For the purpose of a fast calculation within QCD analyses for PDF determination, $\textsc{DiffTop}$ has been interfaced to 
\textsc{fastNLO}~\cite{Britzger:2012bs,Wobisch:2011ij,Kluge:2006xs} which allows the user to 
calculate fast theory predictions using any PDFs. 
\textsc{DiffTop} and its interface to \textsc{fastNLO} provide a framework 
for the inclusion of differential $t\bar{t}$ cross sections at 
approximate NNLO into QCD analyses of PDFs, where a simultaneous determination 
of gluon, $\alpha_s(M_Z)$ and top-quark mass, using the $t\bar{t}$ measurements 
together with measurements of Deep-Inelastic Scattering (DIS) 
and jet production in DIS and proton-proton collisions, might resolve the correlations among these quantities.

\begin{figure}[ht]
\begin{center}
\includegraphics[width=6cm, angle=0]{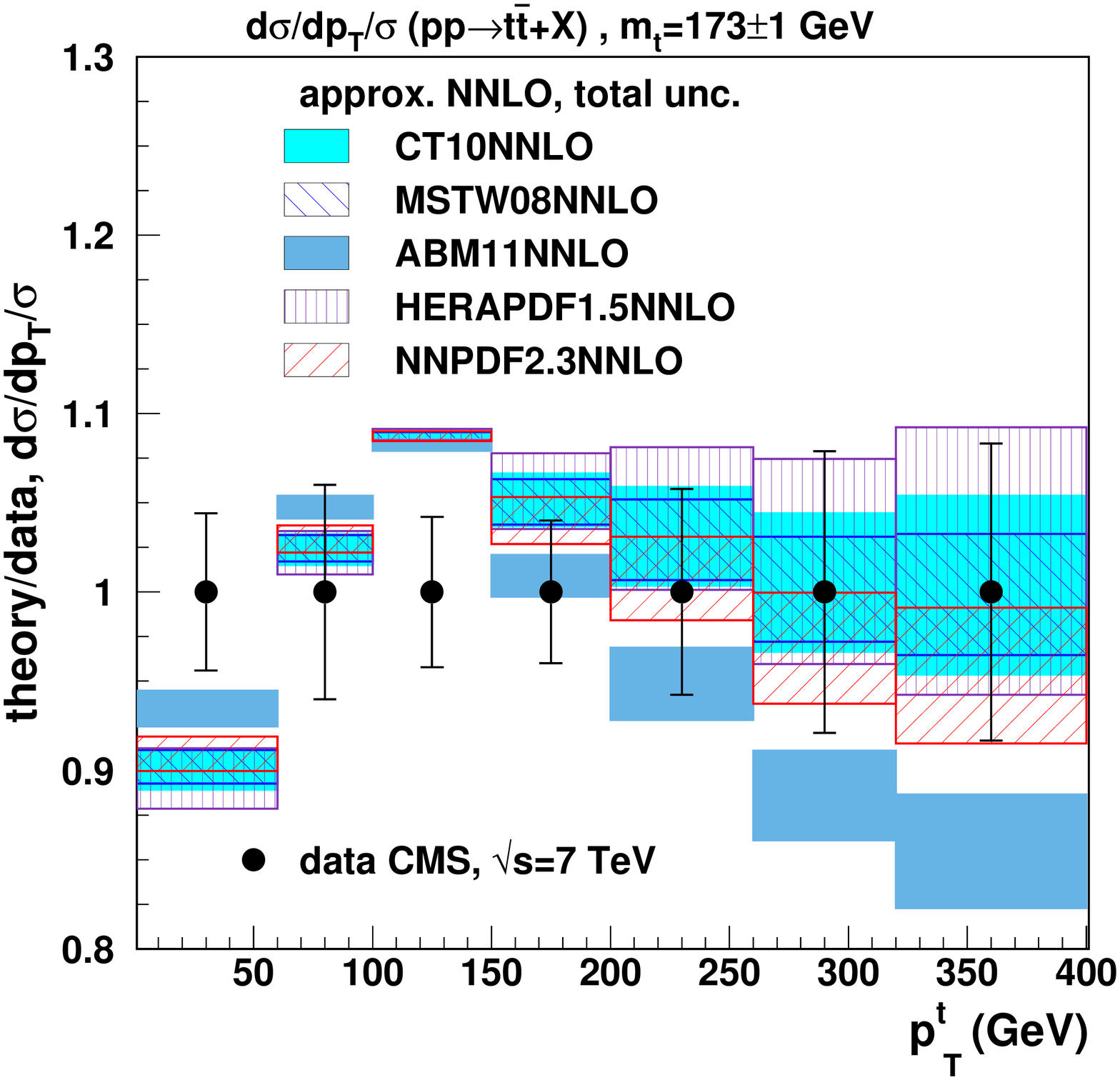}
\includegraphics[width=6cm, angle=0]{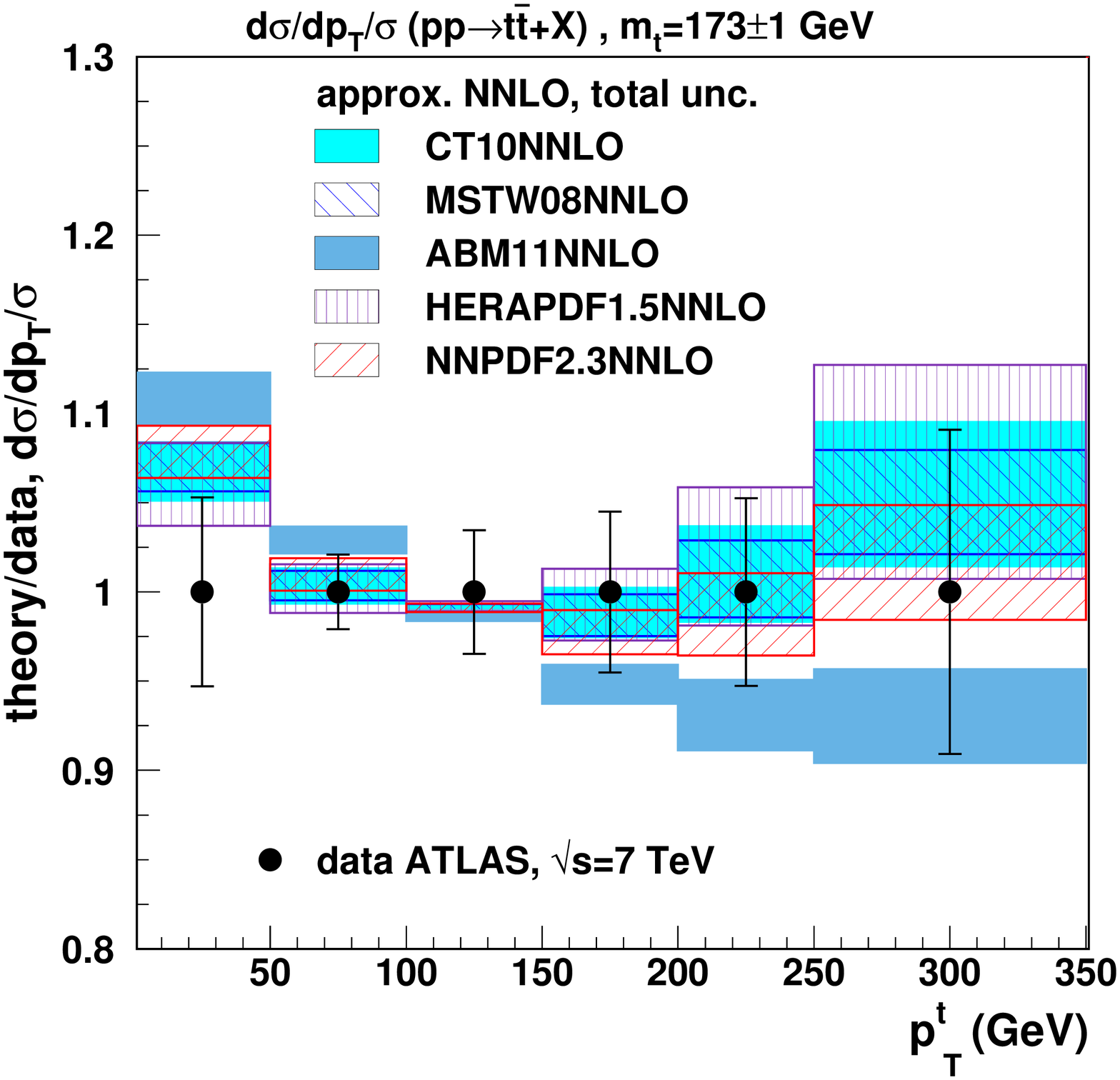}
\hspace{7.5cm}
\caption{The approx. NNLO predictions for top-quark pair production at the LHC at $\sqrt{S} = 7$ TeV, shown as 
functions of $p^t_T$. The predictions, obtained by using different PDF sets are presented as a ratio 
to the LHC measurements (filled symbols). Predictions, obtained by using different PDF sets are 
presented by bands of different hatches. The total uncertainty is obtained by summing the uncertainties due to 
PDFs, $\alpha_s$, $m_t$ and scale variations in quadrature.
\label{allPDFs-unc-Th_dta}}
\end{center}
\end{figure}

{\bf Results.} 
In this section we show results for the approximate NNLO calculation for 
differential cross sections in the single-particle inclusive (1PI) kinematic at the LHC 
(see Refs.\cite{Laenen:1998qw,Kidonakis:2001nj} for details on the kinematics). 
Theory predictions are compared to the recent LHC measurements of differential distributions 
for $t\bar{t}$ production at $\sqrt{S}=7$ TeV by the CMS~\cite{Chatrchyan:2012saa} and ATLAS~\cite{TheATLAScollaboration:2013eja} 
collaborations. In particular, transverse momentum $p^t_T$ distributions for the final-state top quark are presented here.
The central prediction corresponding to the approximate NNLO 
is obtained by choosing $m_t=173$ (pole) GeV, and renormalization 
and factorization scales as $\mu_R=\mu_F=m_t$.
The theory predictions shown for the various PDF sets 
ABM11~\cite{Alekhin:2012ig}, CT10~\cite{Gao:2013xoa}, HERAPDF1.5~\cite{CooperSarkar:2011aa}, 
MSTW08~\cite{Martin:2009iq}, and NNPDF2.3~\cite{Ball:2012cx}, use the $\alpha_s(M_Z)$ value given by each group.
In Fig.~\ref{mstw08-nlo-scale-unc} we compare \textsc{DiffTop} 
and the full NLO calculation obtained by \textsc{MCFM}~\cite{Campbell:2000bg}.
The inclusion of higher orders reduces substantially the scale dependence and also modifies the shape of the $p^t_T$ distribution.
When these theory predictions are compared to the recent LHC measurements 
the theoretical description of the measurements significantly improves in the NNLO case. 
In Fig.~\ref{allPDFs-unc-Th_dta} we compare predictions using different PDF sets to the CMS and ATLAS data.
The errors corresponding to the different PDF sets are represented by bands with different hatches and are estimated by 
summing in quadrature the uncertainties relative to PDF, $\alpha_s(M_Z)$, scale, and $m_t$ variations.
Here PDF uncertainties are shown at 68\% confidence level (CL) 
and are computed by following the prescription given by each PDF group with the exception 
of ABM, in which the total uncertainty, obtained with the symmetric formula for the eigenvector sets, 
represents the PDF + $\alpha_s$ uncertainty at the 68\% CL.
For the MSTW08, CT10, HERAPDF1.5, and NNPDF2.3 PDF sets 
the uncertainty associated to $\alpha_s(M_Z)$ is given by the central
value as given by each group $\pm\Delta\alpha_s(M_Z)= 0.001$.
The scale uncertainty is obtained by variations $m_t/2 \leq \mu_R = \mu_F \leq 2m_t$, while 
the uncertainty associated to the top-quark mass is estimated by
using $m_t= 173$ GeV (pole mass) $\pm\Delta m_t = 1$ GeV.

At the present stage, even though the CMS and ATLAS measurements exhibit relatively large uncertainties, 
these data might have some impact in PDFs determination once included in QCD fit analyses. 
On the other hand, a significant amount of information contained in the measurements of differential distributions 
is lost by normalizing the data. Measurements of absolute differential cross sections are of crucial 
importance to fully exploit the potential of the $t\bar{t}$ production to constrain the gluon distribution.
Moreover, a reduction of the statistic and systematic uncertainties in the high-energy run of the LHC will be of 
clear advantage.

{\bf Conclusions.}
We have shown results for $t\bar{t}$ differential cross sections
obtained with the flexible computer code \textsc{DiffTop} at approximate NNLO, 
which are relevant for phenomenological investigations at the hadron colliders.
We have illustrated theoretical predictions for 
PDF sets that account for uncertainties due to variations of PDFs, scale, $\alpha_s$ and $m_t$,
and that are compared to the recent measurements by the CMS and ATLAS collaborations.
Given the accuracy of the present data and the existing correlations between the strong 
coupling $\alpha_s(M_Z)$, top-quark mass, and gluon PDF,
these measurements might have an impact in constraining the large-$x$ gluon distribution 
once included in QCD fit of PDFs. In particular, investigations of absolute differential cross 
sections will bring complementary information related to the magnitude and other details of the distributions, 
which will be crucial to improve the constraining power of the experimental data.

{\bf Acknowledgments.} This work was supported by the ``Initiative and Networking Fund of 
the Helmholtz Association (HGF) under the contract S0-072''.

\bibliographystyle{h-elsevier3}


\end{document}